\documentclass[12pt]{article}
\usepackage{amsmath,amssymb,amsthm,amsxtra,overpic,bbm,bm,epsfig,subfigure}
\usepackage{hyperref}
\usepackage{mathrsfs}
\usepackage{enumitem}
\usepackage{graphicx}
\usepackage{color}
\usepackage{comment}
\usepackage{epstopdf}
\usepackage{float}
\usepackage{cite}
\usepackage{slashed,stmaryrd}

\usepackage{multirow}
\newcommand{\Dlr}{\mbox{$\raisebox{2mm}{\boldmath ${}^\leftrightarrow$}\hspace{-4mm}D^{}_\mu$}}
\newcommand{\Dilr}{\mbox{$\raisebox{2mm}{\boldmath ${}^\leftrightarrow$}\hspace{-4mm}D^I_\mu$}}
\newcommand{\Dl}{\mbox{$\raisebox{2mm}{\boldmath ${}^\leftarrow$}\hspace{-4mm}D^{}_\mu$}}
\newcommand{\Dls}{\mbox{$\raisebox{2mm}{\boldmath ${}^\leftarrow$}\hspace{-4mm}\slashed{D}$}}
\newcommand{\Dlrn}{\mbox{$\raisebox{2mm}{\boldmath ${}^\leftrightarrow$}\hspace{-4mm}D^\nu$}}
\newcommand{\Dilrn}{\mbox{$\raisebox{2mm}{\boldmath ${}^\leftrightarrow$}\hspace{-4mm}D^{I\nu}$}}

\usepackage{ulem}

\begin{document}

\begin{center}
{\Large\bf Radiative Decays of Charged Leptons in the Seesaw Effective Field Theory with One-loop Matching}
\end{center}

\vspace{0.2cm}

\begin{center}
{\bf Di Zhang~$^{a,~b}$}~\footnote{E-mail: zhangdi@ihep.ac.cn},
\quad
{\bf Shun Zhou~$^{a,~b}$}~\footnote{E-mail: zhoush@ihep.ac.cn (corresponding author)}
\\
\vspace{0.2cm}
{\small
$^a$Institute of High Energy Physics, Chinese Academy of Sciences, Beijing 100049, China\\
$^b$School of Physical Sciences, University of Chinese Academy of Sciences, Beijing 100049, China}
\end{center}

\vspace{1.5cm}

\begin{abstract}
The canonical type-I seesaw model with three heavy Majorana neutrinos is one of the most natural extensions of the standard model (SM) to accommodate tiny Majorana masses of three ordinary neutrinos. At low-energy scales, Majorana neutrino masses and unitarity violation of lepton flavor mixing have been extensively discussed in the literature, which are respectively generated by the unique dimension-five Weinberg operator and one dimension-six operator in the seesaw effective field theory (SEFT) with the tree-level matching. In this work, we clarify that a self-consistent calculation of radiative decays of charged leptons $\beta^- \to \alpha^- + \gamma$ requires the SEFT with one-loop matching, where new six-dimensional operators emerge and make important contributions. For the first time, the Wilson coefficients of all the relevant six-dimensional operators are computed by carrying out the one-loop matching between the effective theory and full seesaw model, and applied to calculate the total rates of radiative decays of charged leptons.
\end{abstract}


\newpage

\framebox{\bf 1} --- In past few decades, a number of elegant neutrino oscillation experiments have firmly established that neutrinos are massive and lepton flavor mixing is significant~\cite{Xing:2019vks}. Obviously, the origin of neutrino masses and lepton flavor mixing calls for new physics beyond the standard model (SM). Among various new-physics extensions of the SM, the canonical type-I seesaw model with three heavy Majorana neutrinos is the simplest and most natural one to accommodate tiny neutrino masses~\cite{Minkowski:1977sc, Yanagida:1979as, GellMann:1980vs, Glashow:1979nm, Mohapatra:1979ia}. Therefore, it is important and necessary to explore the phenomenological consequences of the seesaw model in all aspects, and confront the theoretical predictions with experimental observations. However, the experimental tests of the seesaw model depend very much on the absolute masses of heavy Majorana neutrinos and their interaction strength with the SM particles. In the original version of type-I seesaw model, heavy Majorana neutrino masses are supposed to be close to the energy scale of grand unified theories (GUT), namely, $\Lambda^{}_{\rm GUT} = 2\times 10^{16}~{\rm GeV}$. Although such heavy Majorana neutrinos are impossible to be directly observed in the terrestrial experiments, they can play a crucial role in the early Universe, offering an intriguing explanation for cosmological matter-antimatter asymmetry via the leptogenesis mechanism~\cite{Fukugita:1986hr}. At low-energy scales, it is in practice more convenient to construct an effective theory of the seesaw model, which will be henceforth called Seesaw Effective Field Theory (SEFT), such that the impact of heavy Majorana neutrinos on low-energy physics is incorporated into a series of field operators of mass dimensions higher than four. It is evident that a consistent derivation of higher-dimensional operators is indispensable for us to extract correct information on the seesaw model from low-energy experimental data.

As pointed out by Weinberg long time ago~\cite{Weinberg:1979sa}, if the SM is viewed as an effective theory at the electroweak scale and higher-dimensional operators are included, there exists one unique dimension-five operator ${\cal O}^{(5)} = \overline{\ell^{}_{{\rm L}}} \widetilde{H} \widetilde{H}^{\rm T} \ell^{\rm c}_{{\rm L}}$ that leads to the Majorana mass term of light neutrinos, where $\ell^{}_{\rm L}$ and $H$ are respectively the left-handed lepton doublet and the Higgs doublet, and $\ell^{\rm c}_{{\rm L}} \equiv {\cal C}\overline{\ell^{}_{\rm L}}^{\rm T}$ and $\widetilde{H} \equiv {\rm i}\tau^2 H^*$ with $\tau^2$ being the second Pauli matrix have been defined in the usual way. In general the gauge-invariant Lagrangian of the standard model effective field theory (SMEFT)~\cite{Buchmuller:1985jz,Grzadkowski:2010es} (see, e.g., Ref.~\cite{Brivio:2017vri}, for a recent review) can be written as
\begin{eqnarray}\label{eq:Left}
\mathcal{L}^{}_{\rm SMEFT} = \mathcal{L}^{}_{\rm SM} + \sum_d \mathcal{L}^{(d)} = \mathcal{L}^{}_{\rm SM} + \sum^{n^{}_d}_{i=1} \frac{1}{\Lambda^{d-4}} C^{(d)}_i \mathcal{O}^{(d)}_i \; ,
\end{eqnarray}
where $d > 4$ stands for the mass dimension of the operator ${\cal O}^{(d)}_i$ (for $i = 1, 2, \cdots, n^{}_d$ with $n^{}_d$ being the total number of $d$-dimensional operators), $C^{(d)}_i$ denotes the corresponding Wilson coefficient, and $\Lambda$ is the cutoff scale that can be identified with heavy Majorana neutrino masses in the SEFT. The complete sets of the operators up to a given mass dimension, such as dim-6~\cite{Buchmuller:1985jz, Grzadkowski:2010es}, dim-7~\cite{Lehman:2014jma, Liao:2016hru}, dim-8~\cite{Li:2020gnx, Murphy:2020rsh} and dim-9~\cite{Li:2020xlh,Liao:2020jmn}, in the SMEFT or its extension with sterile neutrinos~\cite{Aparici:2009fh, delAguila:2008ir, Bhattacharya:2015vja, Liao:2016qyd} have been constructed and the one-loop renormalization-group (RG) equations for the dim-5~\cite{Antusch:2001ck}, dim-6~\cite{Jenkins:2013zja,Jenkins:2013wua,Alonso:2013hga} and dim-7~\cite{Liao:2019tep} operators have been derived. On the other hand, for a specific renormalizable model at a high-energy scale, one can match it to the (SM)EFT to study its low-energy consequences. There are only few available examples of complete or partial one-loop matching for the SM extended with a charged scalar singlet~\cite{Bilenky:1993bt}, a real scalar singlet~\cite{Boggia:2016asg, Ellis:2017jns, Jiang:2018pbd, Haisch:2020ahr, Cohen:2020fcu}, a real scalar triplet~\cite{Henning:2016lyp, Ellis:2016enq, Fuentes-Martin:2016uol}, a vector-like quark singlet~\cite{delAguila:2016zcb}, a light sterile neutrino and heavy fermions and a scalar singlet~\cite{Chala:2020vqp}, and two scalar leptoquarks~\cite{Gherardi:2020det} by either diagrammatic calculations or functional approaches~\cite{Henning:2014wua, Drozd:2015rsp, Ellis:2016enq, Henning:2016lyp, Fuentes-Martin:2016uol, Zhang:2016pja, Ellis:2017jns, Kramer:2019fwz, Cohen:2019btp, Cohen:2020fcu, Cohen:2020qvb, Fuentes-Martin:2020udw}. Though some attempts~\cite{Broncano:2002rw, Broncano:2003fq, Broncano:2004tz, Cirigliano:2005ck, Antusch:2006vwa, Abada:2007ux, deGouvea:2007qla, delAguila:2008ir, Gavela:2008ra, Gavela:2009cd, Bonnet:2009ej, Bonnet:2012kz, delAguila:2012nu, Angel:2012ug, Antusch:2014woa, Bhattacharya:2015vja, Davidson:2018zuo, Coy:2018bxr, Barducci:2020ncz, Gargalionis:2020xvt} have also been made to examine the various aspects of the seesaw phenomenology by using EFT techniques,\footnote{The EFT techniques have been widely applied in the flavor physics of heavy quarks. See, e.g., Refs.~\cite{Inami:1980fz,Grinstein:1988me,Lim:1988yu,Buras:1993xp,Buras:1994dj,Malkawi:1995dm,DAmbrosio:2002vsn,Gorbahn:2004my,AguilarSaavedra:2004wm,AguilarSaavedra:2008zc,Calibbi:2015kma,Boos:2020kqq}, for the discussions about the flavor-changing neutral-current processes in the weak decays of heavy quarks.} to our best knowledge, the SEFT with one-loop matching has never been accomplished even in a specific case. As far as the tree-level matching between the SEFT with operators of $d\leq 6$ and the seesaw model is concerned, apart from the five-dimensional Weinberg operator, one six-dimensional operator ${\cal O}^{(6)} = (\overline{\ell^{}_{{\rm L}}} \widetilde{H}) {\rm i} \slashed{\partial} ( \widetilde{H}^\dagger \ell^{}_{{\rm L}} )$ should be taken into account~\cite{Broncano:2002rw, Broncano:2003fq}. It is well known that the Weinberg operator is responsible for neutrino masses and lepton flavor mixing, whereas such a dimension-six operator actually induces the unitarity violation of the lepton flavor mixing matrix~\cite{Antusch:2006vwa}.

In this work, we initiate the complete one-loop matching for the SEFT up to dimension-six operators, and take the radiative decays of charged leptons $\beta^- \to \alpha^- + \gamma$, where $(\alpha, \beta)$ runs over $(e, \mu)$, $(e, \tau)$ and $(\mu, \tau)$, as a well-motivated example to demonstrate the diagrammatic prescription. The complete one-loop matching for SEFT will be done by both diagrammatic calculations and functional approaches in the forthcoming separate works~\cite{Zhang2021}. The experimental searches for radiative $\beta^- \to \alpha^- + \gamma$ decays have placed the ever most stringent bound on lepton flavor violation in the charged-lepton sector, and the discovery of such rare decays will definitely point to new physics beyond the SM~\cite{Zyla:2020zbs, Lindner:2016bgg, Calibbi:2017uvl}. The null signal of  $\beta^- \to \alpha^- + \gamma$ decays has been used to constrain the unitarity violation of Pontecorvo-Maki-Nakagawa-Sakata (PMNS) mixing matrix~\cite{Pontecorvo:1957cp, Maki:1962mu, Pontecorvo:1967fh} in both the minimal unitarity violation (MUV) scheme (with only one tree-level dim-6 operator besides the unique dim-5 operator)~\cite{Antusch:2006vwa, Antusch:2014woa} and the type-I seesaw model~\cite{Fernandez-Martinez:2016lgt, Xing:2020ivm}. The discrepancy between the constraints in these two cases has been noticed~\cite{Fernandez-Martinez:2016lgt, Xing:2020ivm}. In fact, one has to perform the one-loop matching and take into account new dimension-six operators emerging only at the one-loop level, as will be shown and clarified in this work. For this purpose, we put forward a practically efficient method to derive a complete set of higher-dimensional operators relevant for a given physical process and determine the associated Wilson coefficients. Although the general principles for the construction of effective theories can be found in the literature, technical difficulties usually show up case by case.

Strictly speaking, with the one-loop matching, one has to derive the RG equations of the Wilson coefficients at the two-loop level in the effective theories and evolve all physical parameters from the matching scale to the characteristic scale of relevant experiments. After this self-consistent procedure is accomplished, we obtain both the complete set of operators and the associated Wilson coefficients, which can thus be used to calculate the physical observables. However, the derivation of two-loop RG equations will be more involved and beyond the scope of the present work. A dedicated study of this issue is left for future works.

\vspace{0.3cm}

\framebox{\bf 2} --- In order to derive the operators up to dimension-six at the one-loop level, we first write down the Lagrangian of the full theory, i.e., the type-I seesaw model
\begin{eqnarray}\label{eq:Lfull}
\mathcal{L}^{}_{\rm full} &=& \mathcal{L}_{\rm SM} + \overline{N^{}_{\rm R}} {\rm i} \slashed{\partial} N^{}_{\rm R} - \left( \frac{1}{2} \overline{N^{\rm c}_{\rm R}} M N^{}_{\rm R} + \overline{\ell^{}_{\rm L}} Y^{}_\nu \widetilde{H} N^{}_{\rm R} + {\rm h.c.} \right) \; ,
\end{eqnarray}
where $N^{}_{i {\rm R}}$ (for $i= 1,2,3$) are three right-handed neutrino singlets. Without loss of generality, we work in the flavor basis where both the charged-lepton Yukawa coupling matrix $Y^{}_{l}$ and the Majorana mass matrix $M = {\rm Diag}\{M^{}_1, M^{}_2, M^{}_3\}$ of right-handed neutrinos are diagonal. With the tree-level matching~\cite{Broncano:2002rw, Broncano:2003fq}, it is straightforward to obtain the Wilson coefficients of the Weinberg operator ${\cal O}^{(5)}_{\alpha \beta} = \overline{\ell^{}_{\alpha {\rm L}}} \widetilde{H} \widetilde{H}^{\rm T} \ell^{\rm c}_{\beta {\rm L}}$ and the dimension-six operator ${\cal O}^{(6)}_{\alpha \beta} = (\overline{\ell^{}_{\alpha {\rm L}}} \widetilde{H}) {\rm i} \slashed{\partial} ( \widetilde{H}^\dagger \ell^{}_{\beta {\rm L}})$, namely, $C^{(5)}_{\alpha \beta} = \Lambda (Y^{}_\nu M^{-1} Y^{\rm T}_\nu)^{}_{\alpha \beta}/2$ and $C^{(6)}_{\alpha \beta} = \Lambda^2 (Y^{}_\nu M^{-2} Y^\dagger_\nu)^{}_{\alpha \beta}$, where the flavor indices have been explicitly shown. Notice that this dimension-six operator appearing at the tree level is related to two dimension-six operators in the Warsaw basis~\cite{Grzadkowski:2010es} by
\begin{eqnarray}\label{eq:6tree}
(\overline{\ell^{}_{\alpha {\rm L}}} \widetilde{H}) {\rm i} \slashed{\partial} ( \widetilde{H}^\dagger \ell^{}_{\beta {\rm L}}) = \frac{1}{4} \left[ \left(\overline{\ell^{}_{\alpha{\rm L}}} \gamma^\mu \ell^{}_{\beta{\rm L}} \right) \left( H^\dagger {\rm i}  \hspace{0.3mm}\raisebox{2mm}{\boldmath ${}^\leftrightarrow$}\hspace{-4mm}D^{}_\mu H \right) - \left(\overline{\ell^{}_{\alpha{\rm L}}} \gamma^\mu \tau^I \ell^{}_{\beta{\rm L}} \right) \left( H^\dagger {\rm i}  \hspace{0.3mm}\raisebox{2mm}{\boldmath ${}^\leftrightarrow$}\hspace{-4mm}D^I_\mu H \right) \right] \;,
\end{eqnarray}
in which $ \Dlr \equiv D^{}_\mu - \Dl$ and $\Dilr \equiv \tau^I D^{}_\mu - \Dl \tau^I$ with $\tau^I$ (for $I=1,2,3$) being the Pauli matrices and $\Dl$ acting on the left. The  covariant derivative $D^{}_\mu$ is defined as $D^{}_\mu \equiv \partial^{}_\mu -{\rm i} g^{}_1 Y B^{}_\mu - {\rm i} g^{}_2 T^I W^I_{\mu}$ where $Y$ and $T^I$ (for $I=1,2,3$) are the generators of the SM gauge groups $\rm U(1)^{}_Y$ and $\rm SU(2)_L$, respectively, while $g^{}_1$ and $g^{}_2$ are the corresponding gauge coupling constants. This dimension-six operator will modify the couplings of neutrinos with both $W$ and $Z$ gauge bosons, resulting in the unitarity violation of the PMNS matrix.

We focus on the radiative decays of charged leptons $\beta^- \to \alpha^- + \gamma$, which are lepton-flavor-violating and definitely occur in nature in light of nonzero neutrino masses. On the one hand, the tree-level SEFT Lagrangian (i.e., the SM Lagrangian together with the operators ${\cal O}^{(5)}_{\alpha \beta}$ and ${\cal O}^{(6)}_{\alpha \beta}$ arising from the tree-level matching) contributes to the radiative decays of charged leptons via one-loop diagrams mediated by massive neutrinos. On the other hand, among all the dimension-six operators in the Warsaw basis~\cite{Grzadkowski:2010es}, only two operators $\mathcal{O}^{}_{eB,\alpha\beta} = \left(\overline{\ell^{}_{\alpha{\rm L}}} \sigma^{}_{\mu \nu} E^{}_{\beta{\rm R}} \right) H B^{\mu\nu}$ and $\mathcal{O}^{}_{eW,\alpha\beta} = \left(\overline{\ell^{}_{\alpha{\rm L}}} \sigma^{}_{\mu \nu} E^{}_{\beta{\rm R}} \right) \tau^I H W^{I \mu\nu}$ make direct contributions at one-loop level. Therefore, we have to calculate the Wilson coefficients of those two operators via one-loop matching. The basic strategy to achieve this goal is outlined as below.
\begin{itemize}
\item First, one should compute relevant one-light-particle-irreducible diagrams at the one-loop level, and perform the matching between the SEFT and the full theory, where all the external particles are taken to be off-shell. In this matching process, one can obtain the operators in a redundant basis, the so-called Green's basis~\cite{Jiang:2018pbd} (see also Ref.~\cite{Gherardi:2020det}), after applying algebraic, Fierz identities and integration by parts.

\item Second, further making use of equations of motion (EOMs), one may convert the redundant operators in the Green's basis to the independent operators in the Warsaw basis, and thus get the corresponding Wilson coefficients for the operators in the Warsaw basis.
\end{itemize}

In practice, to calculate the Wilson coefficients of $\mathcal{O}^{}_{eB,{\alpha \beta}}$ and $\mathcal{O}^{}_{eW, {\alpha \beta}}$ relevant to the radiative decays of charged leptons, one needs to find out the operators and the associated Wilson coefficients in the Green's basis, which can be converted into $\mathcal{O}^{}_{eB,{\alpha \beta}}$ and $\mathcal{O}^{}_{eW, {\alpha \beta}}$ by applying algebraic, Fierz identities, integration by parts and EOMs. Based on the complete list of dimension-six operators in Ref.~\cite{Grzadkowski:2010es}, we find that those in classes $H^2D^4$, $\psi^2D^3$, $X^2D^2$, $\psi^2 H D^2$, $\psi^2 X D$ and $\psi^2 X H$ (here $\psi$, $X$ and $D$ stand for fermion fields, gauge field strength tensors and covariant derivative, respectively) are relevant. Then, with the help of the package BasisGen~\cite{Criado:2019ugp}, we generate a series of dimension-six operators in these classes, where the redundancies due to algebraic or Fierz transformations or integration by parts have been removed but EOMs have not been applied. In this basis, the operators of our interest are listed in Table~\ref{tb:basis}. Some helpful comments are in order.
\begin{table}[t]
\centering
\renewcommand\arraystretch{1.2}
\begin{tabular}{c|c}
	\hline\hline
	Classes & Operators
	\\\hline
	$H^2D^4$ & $\mathcal{O}^{}_{H} = \left( D^2 H\right)^\dagger \left( D^2 H\right)$
	\\\hline
	$\psi^2D^3$ & $\bm{ \mathcal{O}^{}_{\ell D,\alpha\beta} = \displaystyle \frac{{\rm i}}{2} \overline{\ell^{}_{\alpha{\rm L}}} \left(D^2 \slashed{D} + \slashed{D} D^2 \right) \ell^{}_{\beta {\rm L}} }$ \;,\; $\mathcal{O}^{}_{eD,\alpha\beta} = \displaystyle \frac{\rm i}{2} \overline{E^{}_{\alpha{\rm R}}} \left( D^2 \slashed{D} + \slashed{D} D^2 \right) E^{}_{\beta {\rm R}}$
	\\\hline
	$X^2 D^2$ & $\mathcal{O}^{}_{BD} = D^\rho B^{\mu\nu} D^{}_\rho B^{}_{\mu\nu}$ \;,\; $\mathcal{O}^{}_{WD} = \left( D^\rho W^{\mu\nu} \right)^I  \left(D^{}_\rho W^{}_{\mu\nu}\right)^I $
	\\\hline
	\multirow{2}{*}{$\psi^2HD^2$} & $\bm{ \mathcal{O}^{[1]}_{eH,\alpha\beta} = \overline{\ell^{}_{\alpha{\rm L}}} E^{}_{\beta{\rm R}} D^2 H }$ \;,\; $\bm{ \mathcal{O}^{[2]}_{eH,\alpha\beta} = \overline{\ell^{}_{\alpha{\rm L}}} D^\mu E^{}_{\beta{\rm R}} D^{}_\mu H }$
	\\
	& $\bm{ \mathcal{O}^{[3]}_{eH,\alpha\beta} = \overline{\ell^{}_{\alpha{\rm L}}} D^2 E^{}_{\beta{\rm R}} H }$ \;,\; $\bm{ \mathcal{O}^{[4]}_{eH,\alpha\beta} = {\rm i} \overline{\ell^{}_{\alpha{\rm L}}} \sigma^{\mu\nu} D^{}_\mu E^{}_{\beta{\rm R}} D^{}_\nu H }$
	\\\hline
	\multirow{5}{*}{$\psi^2 X D$} & $\mathcal{O}^{[1]}_{\ell BD,\alpha\beta} = \overline{\ell^{}_{\alpha{\rm L}}} \gamma^\mu \ell^{}_{\beta {\rm L}} D^\nu B^{}_{\mu\nu}$ \;,\; $\bm{ \mathcal{O}^{[2]}_{\ell BD,\alpha\beta} = \displaystyle \frac{\rm i}{2} \overline{\ell^{}_{\alpha{\rm L}}} \gamma^\mu \Dlrn \ell^{}_{\beta {\rm L}} B^{}_{\mu\nu} }$
     \\
     & $\bm{ \mathcal{O}^{[3]}_{\ell BD,\alpha\beta} = \displaystyle \frac{\rm i}{2} \overline{\ell^{}_{\alpha{\rm L}}} \gamma^\mu \Dlrn \ell^{}_{\beta {\rm L}} \widetilde{B}^{}_{\mu\nu} }$ \;,\;  $\mathcal{O}^{[1]}_{eBD,\alpha\beta} = \overline{E^{}_{\alpha{\rm R}}} \gamma^\mu E^{}_{\beta {\rm R}} D^\nu B^{}_{\mu\nu}$
	\\
	& $\mathcal{O}^{[2]}_{eBD,\alpha\beta} = \displaystyle \frac{\rm i}{2} \overline{E^{}_{\alpha{\rm R}}} \gamma^\mu \Dlrn E^{}_{\beta {\rm R}} B^{}_{\mu\nu}$ \;,\; $\mathcal{O}^{[3]}_{eBD,\alpha\beta} = \displaystyle \frac{\rm i}{2} \overline{E^{}_{\alpha{\rm R}}} \gamma^\mu \Dlrn E^{}_{\beta {\rm R}} \widetilde{B}^{}_{\mu\nu}$
	\\
	& $\mathcal{O}^{[1]}_{\ell WD,\alpha\beta} = \overline{\ell^{}_{\alpha{\rm L}}} \gamma^\mu \tau^I \ell^{}_{\beta {\rm L}} \left( D^\nu W^{}_{\mu\nu} \right)^I $ \;,\; $\bm{ \mathcal{O}^{[2]}_{\ell WD,\alpha\beta} = \displaystyle \frac{\rm i}{2} \overline{\ell^{}_{\alpha{\rm L}}} \gamma^\mu \Dilrn  \ell^{}_{\beta {\rm L}} W^I_{\mu\nu}}$
    \\
    & $\bm{ \mathcal{O}^{[3]}_{\ell WD,\alpha\beta} = \displaystyle \frac{\rm i}{2} \overline{\ell^{}_{\alpha{\rm L}}} \gamma^\mu  \Dilrn \ell^{}_{\beta {\rm L}} \widetilde{W}^I_{\mu\nu}}$
	\\\hline
	$\psi^2 X H$ & $\bm{ \mathcal{O}^{}_{eB,\alpha\beta} = \left(\overline{\ell^{}_{\alpha{\rm L}}} \sigma^{}_{\mu \nu} E^{}_{\beta{\rm R}} \right) H B^{\mu\nu} }$ \;,\; $\bm{ \mathcal{O}^{}_{eW,\alpha\beta} = \left(\overline{\ell^{}_{\alpha{\rm L}}} \sigma^{}_{\mu \nu} E^{}_{\beta{\rm R}} \right) \tau^I H W^{I \mu\nu}}$
	\\\hline\hline
\end{tabular}
\vspace{0.8cm}
\caption{The dimension-six operators in classes $H^2D^4$, $\psi^2D^3$, $X^2D^2$, $\psi^2 H D^2$, $\psi^2 X D$ and $\psi^2 X H$, where the Hermitian conjugates of the operators in classes $\psi^2 H D^2$ and $\psi^2 X H$ are not listed explicitly and the most relevant operators for one-loop matching are highlighted in boldface. In addition, $\alpha$ and $\beta$ (for $\alpha,\beta=e,\mu,\tau$) are lepton flavor indices, $D^{}_\rho B^{}_{\mu\nu} \equiv \partial^{}_\rho B^{}_{\mu\nu}$, $\left( D^{}_\rho W^{}_{\mu\nu} \right)^I \equiv \partial^{}_\rho W^I_{\mu\nu} + g^{}_2 \varepsilon^{IJK} W^J_\rho W^K_{\mu \nu}$,  while $\widetilde{X}^{}_{\mu\nu}$ (for $X=B, W^I$) are the dual gauge field strength tensors defined as $\widetilde{X}^{}_{\mu\nu} = \varepsilon^{}_{\mu\nu\rho\sigma} X^{\rho\sigma}/2 $ with $\varepsilon^{}_{0123} = + 1$.}
\label{tb:basis}
\end{table}

\begin{itemize}
\item Since right-handed neutrinos $N^{}_{\rm R}$ only interact with the SM lepton and Higgs doublets via the Yukawa coupling term, the operators $\mathcal{O}^{}_{eD}$, $\mathcal{O}^{}_{JD}$ (for $J=B,W$) and $\mathcal{O}^{[i]}_{eBD}$ (for $i=1,2,3$) only involving right-handed charged leptons $E^{}_{\rm R}$ or gauge bosons (i.e., $B$ and $W$) cannot be generated by right-handed neutrinos at the one-loop level and the corresponding Wilson coefficients are simply vanishing.

\item Even if the EOMs of $H$, $B$ and $W$  are taken into account, the operators $\mathcal{O}^{}_{H}$, $\mathcal{O}^{[1]}_{eH}$ and $\mathcal{O}^{[1]}_{\ell JD}$ (for $J=B, W$) cannot be reduced to $\mathcal{O}^{}_{eJ}$ (for $J=B,W$), implying that they are irrelevant to radiative decays of charged leptons. However, $\mathcal{O}^{[1]}_{eH}$ has to be considered when calculating the Wilson coefficients of other relevant operators, especially those of $\mathcal{O}^{}_{eJ}$ (for $J=B, W$).
\end{itemize}

In summary, only the operators $\mathcal{O}^{}_{\ell D}$, $\mathcal{O}^{[i]}_{eH}$, $\mathcal{O}^{[j]}_{\ell JD}$ and $\mathcal{O}^{}_{eJ}$ (for $i=1,2,3,4$, $j=2,3$ and $J=B, W$), which have been highlighted in boldface in Table~\ref{tb:basis}, need to be considered. Once we obtain their Wilson coefficients, those of $\mathcal{O}^{}_{eJ}$ (for $J=B, W$) in the Warsaw basis can be derived by using algebraic, Fierz identity, integration by parts and the EOMs. This leads us to the complete set of relevant one-loop-level operators with correct Wilson coefficients. Together with the tree-level SEFT Lagrangian, these operators generate the full amplitude of radiative decays of charged leptons, with which the decay rates can be calculated.

\vspace{0.3cm}

\framebox{\bf 3} --- Now we present explicit one-loop matching between the SEFT working in the infrared region (IR) and the full theory in the ultraviolet region (UV), and figure out the Wilson coefficients of relevant operators at one-loop level by using the approach of Feynman diagrams. All the calculations will be performed with the dimensional regularization and the modified minimal subtraction ($\rm \overline{MS}$) scheme for the space-time dimension of ${\sf d} \equiv 4 - 2\epsilon$. The basic idea is to utilize the method of expansion by regions~\cite{Beneke:1997zp, Smirnov:2002pj, Jantzen:2011nz}. More explicitly, for a given one-loop Feynman amplitude in the UV theory, one can split it into hard- (i.e., $k\sim M^{}_i \gg p$ with $k$, $M^{}_i$ and $p$ standing for the loop momentum, heavy particle masses, and external momenta, respectively) and soft-momentum (i.e., $k \sim p \ll M^{}_i$) regions in dimensional regularization. In each region, the integrand of the loop integral is expanded as a Taylor series with respect to the parameters that are considered small, and then the integrand should be integrated over the whole ${\sf d}$-dimensional space of the loop momentum (i.e., $k$). Then, the Wilson coefficients of the operators at one-loop level can be identified by equating the contributions from the hard-momentum region with those from the one-loop operators. The contributions from the soft-momentum region are exactly the same as those from the tree-level SEFT Lagrangian via one-loop diagrams. Consequently, as for one-loop matching, we pay attention only to the hard-momentum region of the amplitudes in the UV and the contributions from the one-loop operators in the IR.
\begin{figure}[t]
  \centering
  \includegraphics[width=0.75\textwidth]{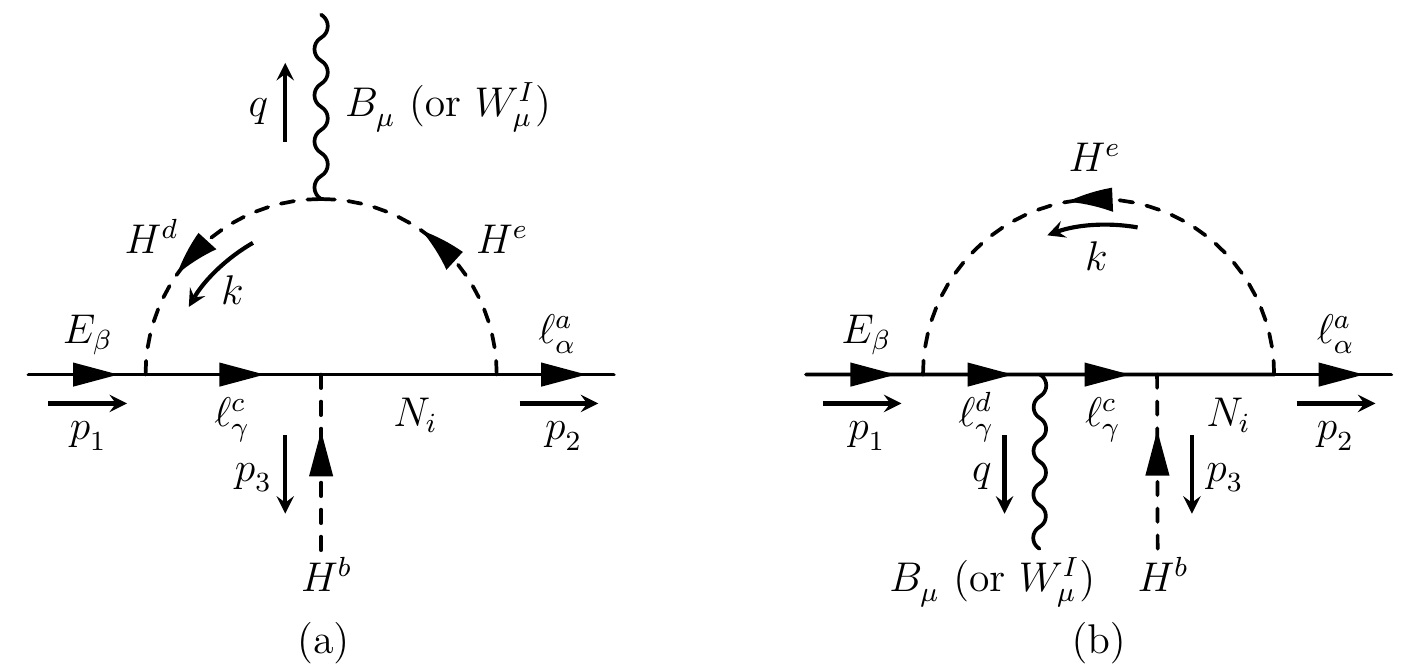}
  \vspace{0.3cm}
  \caption{Feynman diagrams for the amplitude $\langle \ell E H J \rangle$ in the ultraviolet (UV) full theory, which should be matched by the operators $\mathcal{O}^{[i]}_{eH}$ (for $i=1,2,3,4$) and $\mathcal{O}^{}_{eJ}$ (for $J=B, W$) in the infrared (IR) effective theory.}
  \label{fig:LHEX}
\end{figure}

The operators $\mathcal{O}^{[i]}_{eH}$ (for $i=1, 2, 3, 4$) and $\mathcal{O}^{}_{eJ}$ (for $J=B, W$) can be matched by computing the amplitudes given by the Feyman diagrams (a) and (b) in Fig.~\ref{fig:LHEX}. We first consider the one-loop matching for $\mathcal{O}^{[i]}_{eH}$ (for $i=1,2,3,4$) and $\mathcal{O}^{}_{eB}$. The corresponding amplitudes for diagrams (a) and (b) for $B^{}_\mu$ in the UV are
\begin{eqnarray}\label{eq:amp1a}
 {\rm i} \mathcal{M}^{B,\rm UV}_a &=&
 \overline{u} \left(p^{}_2\right) P^{}_{\rm R} \left\{ \int \frac{{\rm d}^4k}{\left(2\pi\right)^4} \frac{\left( \slashed{k}+\slashed{q}+\slashed{p}^{}_2 \right) \left(\slashed{k}+\slashed{p}^{}_1 \right) \left( 2k + q \right)^\mu }{k^2 \left(k+q\right)^2 \left(k+p^{}_1\right)^2 \left[ \left(k+q+p^{}_2\right)^2 - M^2_i\right] } \right\} u\left(p^{}_1\right) \epsilon^\ast_\mu \left(q\right)\nonumber \\
&~& \times \frac{1}{2} g^{}_1 \delta^{ab} \left(Y^{}_\nu\right)^{}_{\alpha i} \left( Y^\dagger_\nu Y^{}_l \right)^{}_{i \beta} \;,
\end{eqnarray}
and
\begin{eqnarray}\label{eq:amp1b}
  {\rm i} \mathcal{M}^{B,\rm UV}_b &=& \overline{u}\left( p^{}_2\right) P^{}_{\rm R} \left\{ \int \frac{{\rm d}^4 k}{\left(2\pi\right)^4} \frac{ \left( \slashed{k} + \slashed{p}^{}_2 \right) \left( \slashed{k} + \slashed{p}^{}_1 - \slashed{q} \right) \gamma^\mu \left( \slashed{k}+\slashed{p}^{}_1 \right) }{k^2 \left( k+p^{}_1 \right)^2 \left( k+p^{}_1-q \right)^2 \left[ \left(k+p^{}_2\right)^2 - M^2_i \right]}  \right\} u\left(p^{}_1\right) \epsilon^\ast_\mu\left(q\right)
  \nonumber
  \\
  && \times \left(-\frac{1}{2}\right) g^{}_1 \delta^{ab} \left( Y^{}_\nu \right)^{}_{\alpha i} \left( Y^\dagger_\nu Y^{}_l \right)^{}_{i \beta} \;,
\end{eqnarray}
where the repeated indices are summed. The hard-momentum parts of these two amplitudes can be obtained by expanding the integrands in the limit of $p \ll k, M^{}_i$ with $p$ being any external momentum (i.e., $p=p^{}_1$, $p^{}_2$, $p^{}_3$, $q$ in this case). Since the combination $\ell E H B$ of relevant fields is already of mass-dimension five, the relevant terms should be proportional to the first power of the external momentum, which is equivalent to one space-time derivative of mass-dimension one. With the help of Eqs.~(\ref{eq:amp1a}) and (\ref{eq:amp1b}), we can get the contributions from the hard-momentum region
\begin{eqnarray}\label{eq:amp1ah}
  {\rm i} \mathcal{M}^{B,{\rm UV}}_a |^{}_{\rm hard} &=& \frac{{\rm i}g^{}_1 \delta^{ab}\left(Y^{}_\nu\right)^{}_{\alpha i} \left( Y^\dagger_\nu Y^{}_l \right)^{}_{i \beta} }{8\left(4\pi\right)^2 M^2_i}  \overline{u} \left(p^{}_2\right) P^{}_{\rm R} \left\{ 2\left[ \gamma^\mu \left( \slashed{p}^{}_1 - \slashed{p}^{}_2 - \slashed{q} \right) - 2\left( p^{}_1 - p^{}_2 - q \right)^\mu \right] \ln\frac{\mu^2}{M^2_i} \right.
  \nonumber   \\
  && + \left. 3\gamma^\mu \left( \slashed{p}^{}_1 - \slashed{p}^{}_2 - \slashed{q} \right) - 2 \left( 3p^{}_1 - 5p^{}_2 -4q \right)^\mu \right\} u\left(p^{}_1\right) \epsilon^\ast_\mu\left(q\right) \; ,
\end{eqnarray}
and
\begin{eqnarray}\label{eq:amp1bh}
  {\rm i} \mathcal{M}^{B,{\rm UV}}_b |^{}_{\rm hard}
  &=& - \frac{{\rm i}g^{}_1 \delta^{ab}\left(Y^{}_\nu\right)^{}_{\alpha i} \left( Y^\dagger_\nu Y^{}_l \right)^{}_{i \beta} }{8\left(4\pi\right)^2 M^2_i}  \overline{u} \left(p^{}_2\right) P^{}_{\rm R} \left\{ 2\left[ \gamma^\mu \left( \slashed{p}^{}_2 + \slashed{q} \right) + 2\left( p^{}_1 - p^{}_2 - q \right)^\mu \right] \ln\frac{\mu^2}{M^2_i} \right.
  \nonumber
  \\
  &&+ \left. \gamma^\mu \left( 3\slashed{p}^{}_2 + \slashed{q} \right) + 2 \left( p^{}_1 - p^{}_2 -q \right)^\mu \right\} u\left(p^{}_1\right) \epsilon^\ast_\mu\left(q\right) \;,
\end{eqnarray}
where only the terms proportional to $p = p^{}_1$, $p^{}_2$, $p^{}_3$, $q$ have been retained and $\mu$ is the renormalization scale. The divergences have not been explicitly shown but they can be easily recovered by setting $\ln (\mu^2/M^2_i) \rightarrow \ln (\mu^2/M^2_i) + \Delta^{}_\epsilon$, where $\Delta^{}_{\epsilon} \equiv 1/\epsilon - \gamma^{}_{\rm E} + \ln \left(4\pi\right)$ with $\gamma^{}_{\rm E}$ being the Euler constant. Then the hard-momentum part of the total amplitude reads
\begin{eqnarray}\label{eq:amp1t}
  {\rm i} \mathcal{M}^{B,{\rm UV}}_{\rm tot} |^{}_{\rm hard} &=& {\rm i} \mathcal{M}^{B,{\rm UV}}_a |^{}_{\rm hard} + {\rm i} \mathcal{M}^{B,{\rm UV}}_b |^{}_{\rm hard}
  \nonumber
  \\
  &=& \frac{{\rm i}g^{}_1 \delta^{ab}\left(Y^{}_\nu\right)^{}_{\alpha i} \left( Y^\dagger_\nu Y^{}_l \right)^{}_{i \beta} }{8\left(4\pi\right)^2 M^2_i}  \overline{u} \left(p^{}_2\right) P^{}_{\rm R} \left\{ 2\left[ \gamma^\mu \left( \slashed{p}^{}_1 - 2\slashed{p}^{}_2 - 2\slashed{q} \right) - 4\left( p^{}_1 - p^{}_2 - q \right)^\mu \right] \ln\frac{\mu^2}{M^2_i} \right.
  \nonumber
  \\
  &&+ \left. \gamma^\mu \left( 3\slashed{p}^{}_1 - 6\slashed{p}^{}_2 - 4\slashed{q} \right) - 2 \left( 4p^{}_1 - 6p^{}_2 -5q \right)^\mu \right\} u\left(p^{}_1\right) \epsilon^\ast_\mu\left(q\right) \;.
\end{eqnarray}
On the other hand, the contribution of one-loop operators $\mathcal{O}^{[i]}_{eH}$ (for $i=1,2,3,4$) and $\mathcal{O}^{}_{eB}$ in the IR to the amplitude $\left\langle \ell E H B \right\rangle$ is given by
\begin{eqnarray}\label{eq:amp1e}
  {\rm i} \mathcal{M}^{B,{\rm EFT}}_{\rm tot} |^{}_{\rm loop}
  &=& \frac{{\rm i}}{2\Lambda^2} g^{}_1 \delta^{ab} \overline{u}\left(p^{}_2\right) P^{}_{\rm R} \left[ \gamma^\mu \slashed{q} \left( \frac{4}{g^{}_1} C^{}_{eB} - 2 C^{[4]}_{eH} \right)^{}_{\alpha\beta} + \gamma^\mu\left(\slashed{p}^{}_1 -2\slashed{p}^{}_2\right) \left( C^{[4]}_{eH} \right)^{}_{\alpha\beta} \right.
  \nonumber
  \\
  && + q^\mu \left( -\frac{4}{g^{}_1} C^{}_{eB} - C^{[1]}_{eH} + 2 C^{[2]}_{eH} - 2 C^{[3]}_{eH} + 2 C^{[4]}_{eH} \right)^{}_{\alpha\beta} + p^\mu_1 \left( 2 C^{[1]}_{eH} - 3 C^{[2]}_{eH} + 4 C^{[3]}_{eH} \right.
  \nonumber
  \\
  &&- \left.\left. C^{[4]}_{eH} \right)^{}_{\alpha\beta} + 2 p^\mu_2 \left( - C^{[1]}_{eH} + C^{[2]}_{eH} + C^{[4]}_{eH} \right)^{}_{\alpha\beta} \right] \;,
\end{eqnarray}
where $C^{[i]}_{eH}$ (for $i=1,2,3,4$) and $C^{}_{eB}$ stand for the Wilson coefficients of $\mathcal{O}^{[i]}_{eH}$ (for $i=1,2,3,4$) and $\mathcal{O}^{}_{eB}$, respectively. Matching at the energy scale $\mu = {\cal O}(M^{}_i)$~\footnote{In the present work, we assume that the masses of heavy Majorana neutrinos are nearly degenerate, namely, $M^{}_1 \approx M^{}_2 \approx M^{}_3$. Therefore, the matching can be done at an average scale $\mu = M^{}_i$ and the RG-running effects between any two mass scales of different heavy Majorana neutrinos can be neglected. If the mass spectrum of heavy Majorana neutrinos is hierarchical, one should integrated them out sequentially, and then the RG-running effects between different mass scales should be taken into account. Furthermore, the matching scale does not have to be exactly at $\mu=M^{}_i$, as emphasized in Ref.~\cite{Antusch:2015pda}. It is usually convenient to choose $\mu = M^{}_i$ in order to avoid the appearance of large logarithms and thus maintain the fast convergence of perturbation calculations.} and dropping the terms proportional to $\Delta^{}_\epsilon$ in Eq.~(\ref{eq:amp1t}), one arrives at an array of five linear equations for five unknown Wilson coefficients by equating Eq.~(\ref{eq:amp1t}) with Eq.~(\ref{eq:amp1e}), namely,
\begin{equation}\label{eq:equ1}
  \left\{
  \begin{array}{rl}
  \displaystyle - \frac{1}{\left(4\pi\right)^2} \cdot S  & = \displaystyle \frac{1}{\Lambda^2} \cdot \left( \frac{4}{g^{}_1} C^{}_{eB} - 2 C^{[4]}_{eH} \right) \;,
  \\
  \displaystyle  \frac{1}{\left(4\pi\right)^2} \cdot \frac{3}{4} S  & = \displaystyle \frac{1}{\Lambda^2} \cdot C^{[4]}_{eH} \;,
  \\
  \displaystyle \frac{1}{\left(4\pi\right)^2} \cdot \frac{5}{2} S  & = \displaystyle \frac{1}{\Lambda^2} \cdot \left(- \frac{4}{g^{}_1} C^{}_{eB} - C^{[1]}_{eH} + 2 C^{[2]}_{eH} - 2 C^{[3]}_{eH} + 2 C^{[4]}_{eH} \right) \;,
  \\
  \displaystyle -\frac{1}{\left(4\pi\right)^2} \cdot 2 S & = \displaystyle \frac{1}{\Lambda^2} \cdot \left(2 C^{[1]}_{eH} - 3 C^{[2]}_{eH} + 4 C^{[3]}_{eH} - C^{[4]}_{eH} \right) \;,
  \\
  \displaystyle \frac{1}{\left(4\pi\right)^2} \cdot 3 S  & = \displaystyle \frac{1}{\Lambda^2} \cdot \left( - 2 C^{[1]}_{eH} + 2 C^{[2]}_{eH} + 2 C^{[4]}_{eH} \right) \;,
  \end{array}
  \right.
\end{equation}
with $S \equiv Y^{}_\nu M^{-2} Y^\dagger_\nu Y^{}_l$. Solving these linear equations in Eq.~(\ref{eq:equ1}), we have
\begin{eqnarray}\label{eq:wc1}
  C^{[1]}_{eH} = \frac{S \Lambda^2}{\left(4\pi\right)^2} \;,\;\; C^{[2]}_{eH} = \frac{7S \Lambda^2}{4\left(4\pi\right)^2} \;,\;\; C^{[3]}_{eH} = \frac{S \Lambda^2}{2\left(4\pi\right)^2} \;,\;\; C^{[4]}_{eH} = \frac{3S \Lambda^2}{4\left(4\pi\right)^2} \;,\;\; C^{}_{eB} = \frac{g^{}_1 S \Lambda^2}{8\left(4\pi\right)^2} \; .
\end{eqnarray}

To obtain the hard-momentum part of the total amplitude $\left\langle \ell E H W \right\rangle$ in Fig.~\ref{fig:LHEX}, one can simply replace $g^{}_1 \delta^{ed}$ with $g^{}_2 \tau^I_{de}$ for diagram (a), and $g^{}_1 \delta^{cd}$ with $-g^{}_2 \tau^I_{cd}$ for diagram (b) in Fig.~\ref{fig:LHEX}, due to the different Feynman rules for the vertices involving $B$ and $W$. Such replacements give
\begin{eqnarray}\label{eq:amp1tw}
   {\rm i} \mathcal{M}^{W,{\rm UV}}_{\rm tot} |^{}_{\rm hard} &=& \frac{-{\rm i}g^{}_2 \tau^I_{ab}}{8\left(4\pi\right)^2 M^2_i}  \left(Y^{}_\nu\right)^{}_{\alpha i} \left( Y^\dagger_\nu Y^{}_l \right)^{}_{i \beta}  \overline{u} \left(p^{}_2\right) P^{}_{\rm R} \left\{ 2\gamma^\mu \slashed{p}^{}_1 \ln\frac{\mu^2}{M^2_i} + \gamma^\mu \left( 3\slashed{p}^{}_1 - 2 \slashed{q} \right) \right.
   \nonumber
   \\
   && - \left. 2 \left( 2p^{}_1 - 4p^{}_2 -3q \right)^\mu \phantom{\frac{1}{1}}\hspace{-0.35cm} \right\} u\left(p^{}_1\right) \epsilon^\ast_\mu\left(q\right) \;.
\end{eqnarray}
Meanwhile, the contribution of one-loop operators $\mathcal{O}^{[i]}_{eH}$ (for $i=1,2,3,4$) and $\mathcal{O}^{}_{eW}$ to the total amplitude $\langle \ell E H W \rangle$ in the IR is given by
\begin{eqnarray}\label{eq:amp1ew}
  {\rm i} \mathcal{M}^{W,{\rm EFT}}_{\rm tot} |^{}_{\rm loop}
  &=& \frac{\rm i}{2\Lambda^2} g^{}_2 \tau^I_{ab} \overline{u} \left(p^{}_2\right) P^{}_{\rm R} \left[ \gamma^\mu \slashed{q} \frac{4}{g^{}_2} \left( C^{}_{eW} \right)^{}_{\alpha\beta} - \gamma^\mu \slashed{p}^{}_1 \left( C^{[4]}_{eH} \right)^{}_{\alpha\beta} - q^\mu \left( \frac{4}{g^{}_2} C^{}_{eW} + C^{[1]}_{eH} \right)^{}_{\alpha\beta} \right.
  \nonumber
  \\
  && + \left. p^\mu_1 \left( 2C^{[1]}_{eH} - C^{[2]}_{eH} + C^{[4]}_{eH} \right)^{}_{\alpha\beta} - 2p^\mu_2 \left( C^{[1]}_{eH} \right)^{}_{\alpha\beta} \right] \;.
\end{eqnarray}
Similarly, by equating Eq.~(\ref{eq:amp1tw}) with Eq.~(\ref{eq:amp1ew}), one can obtain
\begin{eqnarray}\label{eq:equ1w}
  \left\{
  \begin{array}{rl}
    \displaystyle \frac{1}{\left(4\pi\right)^2} \cdot \frac{1}{2} S & = \displaystyle \frac{1}{\Lambda^2} \cdot \frac{4}{g^{}_2} C^{}_{eW} \;,
    \\
    \displaystyle \frac{1}{\left(4\pi\right)^2} \cdot \frac{3}{4} S & = \displaystyle \frac{1}{\Lambda^2} \cdot C^{[4]}_{eH} \;,
    \\
    \displaystyle \frac{1}{\left(4\pi\right)^2} \cdot \frac{3}{2} S & = \displaystyle \frac{1}{\Lambda^2} \cdot \left( \frac{4}{g^{}_2} C^{}_{eW} + C^{[1]}_{eH} \right) \;,
    \\
    \displaystyle \frac{1}{\left(4\pi\right)^2} \cdot S & = \displaystyle \frac{1}{\Lambda^2} \cdot \left( 2C^{[1]}_{eH} - C^{[2]}_{eH} + C^{[4]}_{eH} \right) \;,
    \\
    \displaystyle \frac{1}{\left(4\pi\right)^2} \cdot 2 S & = \displaystyle \frac{1}{\Lambda^2} \cdot 2C^{[1]}_{eH} \; ,
  \end{array}
  \right.
\end{eqnarray}
where one can observe that the solutions to $C^{[i]}_{eH}$ (for $i=1, 2, 4$) are identical to those given in Eq.~(\ref{eq:wc1}) as they should be. In addition, the new Wilson coefficient $C^{}_{eW}$ is found to be
\begin{eqnarray}\label{eq:wc1w}
  C^{}_{eW} = \frac{g^{}_2 S \Lambda^2}{8\left(4\pi\right)^2} \; .
\end{eqnarray}

\begin{figure}[t]
  \centering
  \includegraphics[width=1\textwidth]{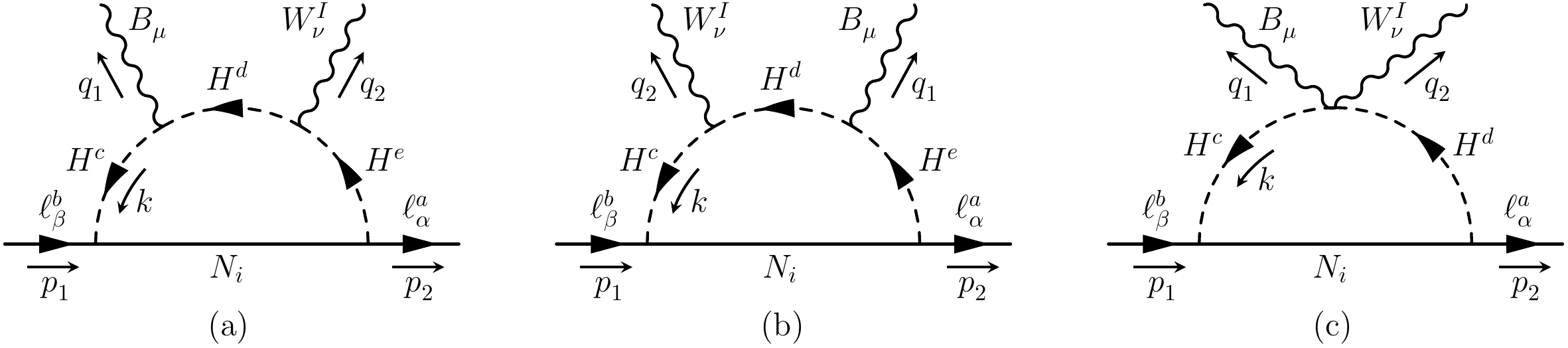}
\vspace{-0.5cm}
  \caption{Feynman diagrams for the amplitude $\langle \ell \ell B W \rangle$ in the UV full theory, which should be matched by the operators $\mathcal{O}^{}_{\ell D}$ and $\mathcal{O}^{[2]}_{\ell JD}$ (for $J = B, W$) in the IR effective theory.}
  \label{fig:LLXX}
\end{figure}

Next we proceed with the one-loop matching of the operators $\mathcal{O}^{}_{\ell D}$ and $\mathcal{O}^{[i]}_{\ell JD}$ (for $i=2,3$ and $J=B,W$) by computing the amplitude $\left\langle \ell \ell B W \right\rangle$ corresponding to three Feynman diagrams in Fig.~\ref{fig:LLXX}. As the techniques for the computations are quite similar to those in the previous case, we just summarize the final result for the hard-momentum part of the total amplitude, i.e.,
\begin{eqnarray}\label{eq:amp2t}
  {\rm i} \mathcal{M}^{\rm UV}_{\rm tot}|^{}_{\rm hard} &=& - \frac{{\rm i} g^{}_1 g^{}_2 \tau^I_{ab}}{12\left(4\pi\right)^2 M^2_i} \left(Y^{}_\nu\right)^{}_{\alpha i} \left(Y^\dagger_\nu\right)^{}_{i\beta} \overline{u}\left(p^{}_2\right) P^{}_{\rm R} \left[ g^{\mu\nu} \left( 2\slashed{p}^{}_1 - \slashed{q}^{}_1 - \slashed{q}^{}_2 \right) + \left( 2p^{}_1 - q^{}_1 - q^{}_2 \right)^\mu \gamma^\nu \right.
  \nonumber
  \\
  && + \left. \left( 2p^{}_1 - q^{}_1 - q^{}_2 \right)^\nu \gamma^\mu \right] u\left(p^{}_1\right) \epsilon^\ast_\mu\left(q^{}_1\right) \epsilon^{I \ast}_\nu\left(q^{}_2\right) \;,
\end{eqnarray}
where the UV divergences completely cancel out. In the IR effective theory, the contributions of the operators $\mathcal{O}^{}_{\ell D}$ and $\mathcal{O}^{[i]}_{\ell JD}$ (for $i=2,3$ and $J=B,W$) are
\begin{eqnarray}\label{eq:amp2e}
  {\rm i} \mathcal{M}^{\rm EFT}_{\rm tot}|^{}_{\rm loop} &=& - \frac{\rm i}{4\Lambda^2} g^{}_1 g^{}_2 \tau^I_{ab} \overline{u} \left( p^{}_2 \right) P^{}_{\rm R} \left\{ \left(C^{}_{\ell D} \right)^{}_{\alpha\beta} \left[ g^{\mu\nu} \left( \slashed{q}^{}_1 + \slashed{q}^{}_2 - 2\slashed{p}^{}_1 \right) + \gamma^\mu \left( q^{}_1 + q^{}_2 - 2p^{}_1 \right)^\nu
  \right.\right.
  \nonumber
  \\
  && + \left. \gamma^\nu \left( q^{}_1 + q^{}_2 - 2p^{}_1 \right)^\mu \phantom{\slashed{p}^{}_1}\hspace{-0.45cm} \right] - \frac{2\rm i}{g^{}_1} \left( C^{[2]}_{\ell B D} \right)^{}_{\alpha\beta} \left( g^{\mu\nu} \slashed{q}^{}_1 - \gamma^\mu q^\nu_1 \right) + \frac{2}{g^{}_1} \left( C^{[3]}_{\ell B D} \right)^{}_{\alpha\beta}
  \nonumber
  \\
  && \times \left[ g^{\mu\nu} \slashed{q}^{}_1 - \gamma^\mu q^\nu_1 + \gamma^\nu \left( q^\mu_1 - \gamma^\mu \slashed{q}^{}_1 \right)  \right] + \frac{2\rm i}{g^{}_2} \left( C^{[2]}_{\ell W D} \right)^{}_{\alpha\beta} \left( g^{\mu\nu} \slashed{q}^{}_2 - \gamma^\nu q^\mu_2 \right)
  \nonumber
  \\
  && \left.  - \frac{2}{g^{}_2}\left( C^{[3]}_{\ell W D} \right)^{}_{\alpha\beta} \left[ g^{\mu\nu} \slashed{q}^{}_2 - \gamma^\nu q^\mu_2 + \gamma^\mu \left( q^\nu_2 - \gamma^\nu \slashed{q}^{}_2 \right)  \right] \right\} u \left( p^{}_1 \right) \epsilon^\ast_\mu\left(q^{}_1\right) \epsilon^{I \ast}_\nu\left(q^{}_2\right) \;,
\end{eqnarray}
which should be identified with Eq.~(\ref{eq:amp2t}), leading to the relevant Wilson coefficients
\begin{eqnarray}\label{eq:wc2}
  C^{}_{\ell D} = - \frac{\Lambda^2}{3\left(4\pi\right)^2} Y^{}_\nu M^{-2} Y^\dagger_\nu \;, \quad  C^{[2]}_{\ell B D} = C^{[3]}_{\ell B D} = C^{[2]}_{\ell W D} = C^{[3]}_{\ell W D} = 0 \;.
\end{eqnarray}

It is worth pointing out that in the IR effective theory, in addition to the contributions from the loop-level operators, the tree-level dimension-six operator ${\cal O}^{(6)}_{\alpha \beta} = (\overline{\ell^{}_{\alpha {\rm L}}} \widetilde{H}) {\rm i} \slashed{\partial} ( \widetilde{H}^\dagger \ell^{}_{\beta {\rm L}})$ also contributes to the amplitudes $\left\langle \ell E H J \right\rangle$ (for $J=B,W$) and $\left\langle \ell \ell B W \right\rangle$ via one-loop diagrams with an $LLHH$ vertex. However, this contribution is identical with the soft-momentum part of the corresponding amplitudes in the UV full theory, which is calculated by expanding the heavy Majorana neutrino propagator ${\rm i}/(\slashed{k} - M^{}_i)$ in terms of $k/M^{}_i$. This expansion converts the nonlocal propagator into local terms and leads to an equivalent $LLHH$ vertex. Thus, these two contributions are irrelevant to the one-loop matching as we have mentioned at the beginning of this section.

Finally, we make use of the EOMs to transform the resultant dimension-six operators into those in the Warsaw basis and identify the corresponding Wilson coefficients of the latter. For radiative decays of charged lepton, only $\mathcal{O}^{}_{eJ}$ (for $J=B,W$) that are contained in the Warsaw basis make direct contributions. Hence we concentrate on the operators $\mathcal{O}^{}_{\ell D}$, $\mathcal{O}^{[i]}_{eH}$ and $\mathcal{O}^{[j]}_{\ell J D}$, which are related to $\mathcal{O}^{}_{eJ}$ (for $i=2,3,4$, $j=2,3$ and $J=B,W$) via EOMs and important for the determination of the Wilson coefficients, while it is unnecessary to consider $\mathcal{O}^{[j]}_{\ell J D}$ (for $j=2,3$ and $J=B,W$) for their Wilson coefficients are vanishing as shown in Eq.~(\ref{eq:wc2}). After some algebraic computations, we find
\begin{eqnarray}
\mathcal{O}^{[2]}_{eH,\alpha\beta} & \Rightarrow & \frac{1}{4} \overline{\ell^{}_{\alpha {\rm L}}}  {\rm i} \sigma^{\mu\nu}  E^{}_{\beta {\rm R}} \left[ D^{}_\mu, D^{}_\nu \right] H =  \frac{1}{8} \left( g^{}_1 \mathcal{O}^{}_{eB,\alpha\beta} + g^{}_2 \mathcal{O}^{}_{eW,\alpha\beta} \right) \; , \label{eq:eh2}\\
\mathcal{O}^{[3]}_{eH,\alpha\beta} &\Rightarrow& \frac{1}{2} \overline{\ell^{}_{\alpha{\rm L}}} {\rm i} \sigma^{\mu\nu} \left[ D^{}_\mu, D^{}_\nu \right] E^{}_{\beta {\rm R}} H =  -\frac{1}{2} g^{}_1 \mathcal{O}^{}_{eB,\alpha\beta} \; , \label{eq:eh3}\\
\mathcal{O}^{[4]}_{eH,\alpha\beta}  & \Rightarrow & - \mathcal{O}^{[2]}_{eH,\alpha\beta} \Rightarrow  - \frac{1}{8} \left( g^{}_1 \mathcal{O}^{}_{eB,\alpha\beta} + g^{}_2 \mathcal{O}^{}_{eW,\alpha\beta} \right) \; , \label{eq:eh4}\\
\mathcal{O}^{}_{\ell D,\alpha\beta} &\Rightarrow&   - \frac{1}{8}  \overline{\ell^{}_{\alpha{\rm L}}} \left[  \sigma^{\mu\nu} \left( g^{}_1 B^{}_{\mu\nu} - g^{}_2 \tau^I W^I_{\mu\nu} \right) {\rm i}\slashed{D}  - {\rm i} \Dls  \sigma^{\mu\nu} \left( g^{}_1 B^{}_{\mu\nu} - g^{}_2 \tau^I W^I_{\mu\nu} \right)\right] \ell^{}_{\beta{\rm L}}
\nonumber
\\
&=& - \frac{1}{8} \left(Y^{}_l\right)^{}_{\beta \gamma} \left(  g^{}_1 \mathcal{O}^{}_{eB,\alpha\gamma} - g^{}_2 \mathcal{O}^{}_{eW,\alpha\gamma} \right) - \frac{1}{8} \left(Y^\dagger_l\right)^{}_{\gamma \alpha} \left(  g^{}_1 \mathcal{O}^{}_{eB,\beta\gamma} - g^{}_2 \mathcal{O}^{}_{eW,\beta\gamma} \right)^\dagger  \;, \label{eq:eh5}
\end{eqnarray}
where the terms irrelevant to the coefficients of $\mathcal{O}^{}_{eJ}$ (for $J=B,W$) have been neglected and the EOM of $\ell^{}_{\rm L}$ has been applied in the last step of Eq.~(\ref{eq:eh5}). According to Eqs.~(\ref{eq:eh2})-(\ref{eq:eh5}), together with the coefficients derived in Eqs.~(\ref{eq:wc1}), (\ref{eq:wc1w}) and (\ref{eq:wc2}), one can obtain the final Wilson coefficients of the operators $\mathcal{O}^{}_{eJ}$ (for $J=B, W$) in the Warsaw basis
\begin{eqnarray}\label{eq:wc}
  C^\prime_{eB} &=& C^{}_{eB} + \frac{g^{}_1}{8}C^{[2]}_{eH} - \frac{g^{}_1}{2} C^{[3]}_{eH} - \frac{g^{}_1}{8} C^{[4]}_{eH} - \frac{g^{}_1}{8} C^{}_{\ell D} Y^{}_l = \frac{g^{}_1 S \Lambda^2}{24\left( 4\pi\right)^2} \;,
  \nonumber
  \\
  C^\prime_{eW} &=& C^{}_{eW} + \frac{g^{}_2}{8}C^{[2]}_{eH} - \frac{g^{}_2}{8} C^{[4]}_{eH} + \frac{g^{}_2} {8} C^{}_{\ell D} Y^{}_l = \frac{5g^{}_2 S \Lambda^2}{24\left(4\pi\right)^2} \;,
\end{eqnarray}
with $S \equiv Y^{}_\nu M^{-2} Y^\dagger_\nu Y^{}_l$. Therefore, the one-loop SEFT Lagrangian with dimension-six operators relevant for radiative decays of charged leptons is given by
\begin{eqnarray}\label{eq:loop}
  \mathcal{L}^{(6)}_{\rm loop} = \frac{\left( Y^{}_\nu M^{-2} Y^\dagger_\nu Y^{}_l \right)^{}_{\alpha\beta}}{24\left(4\pi\right)^2} \left[ g^{}_1 \left(\overline{\ell^{}_{\alpha {\rm L}}} \sigma^{}_{\mu\nu} E^{}_{\beta {\rm R}} \right) H B^{\mu\nu} + 5g^{}_2 \left( \overline{\ell^{}_{\alpha{\rm L}}} \sigma^{}_{\mu\nu} E^{}_{\beta{\rm R}} \right) \tau^I H W^{I \mu\nu} \right] + {\rm h.c.} \; ,
\end{eqnarray}
which will be used to calculate the decay rate for $\beta^- \to \alpha^- + \gamma$ in the next section. It is worth stressing that these loop-level dimension-six operators should be added to the tree-level SEFT Lagrangian with the Weinberg operator and the tree-level dimension-six operator.

\vspace{0.3cm}

\framebox{\bf 4} --- The one-loop SEFT Lagrangian with operators up to dimension-six has been specified, and its explicit form after the spontaneous gauge symmetry breaking reads
\begin{eqnarray}\label{eq:Lssb}
  \mathcal{L}^{}_{\rm SEFT} &=&  \overline{\nu^{}_{\alpha{\rm L}}} \left( \bm{1} + M^{}_{\rm D} M^{-2} M^\dagger_{\rm D} \right)^{}_{\alpha\beta} {\rm i} \slashed{\partial} \nu^{}_{\beta{\rm L}} - \left[ \overline{l^{}_{\alpha{\rm L}}} \left(M^{}_l\right)^{}_{\alpha\beta} l^{}_{\beta{\rm R}} + \frac{1}{2} \overline{\nu^{}_{\alpha{\rm L}}} \left(M^{}_\nu\right)^{}_{\alpha\beta} \nu^{\rm c}_{\beta{\rm L}} + {\rm h.c.} \right]   \nonumber   \\
  && + \left( \frac{g^{}_2}{\sqrt{2}} \overline{l^{}_{\alpha{\rm L}}} \gamma^\mu \nu^{}_{\alpha{\rm L}} W^-_\mu + {\rm h.c.} \right) + \frac{g^{}_2}{2\cos \theta^{}_{\rm w}} \overline{\nu^{}_{\alpha{\rm L}}} \gamma^\mu \nu^{}_{\alpha{\rm L}} Z^{}_\mu
  \nonumber
  \\
  && + \frac{g^2_2 \left( M^{}_{\rm D} M^{-2} M^\dagger_{\rm D} M^{}_l \right)^{}_{\alpha\beta} }{48\left(4\pi\right)^2 M^2_W} \left( g^{}_1 \cos \theta^{}_{\rm w} - 5g^{}_2 \sin \theta^{}_{\rm w}  \right) \overline{l^{}_{\alpha{\rm L}}} \sigma^{}_{\mu\nu} l^{}_{\beta{\rm R}} F^{\mu\nu} + {\rm h.c.} \; ,
\end{eqnarray}
where only the relevant terms for lepton masses, flavor mixing and radiative decays of charged leptons are kept. Some explanations for our notations are helpful. First, we have the charged-lepton mass matrix $M^{}_l \equiv v Y^{}_l/ \sqrt{2}$ with $v \approx 246~{\rm GeV}$ being the vacuum expectation value of the Higgs field, the Dirac neutrino mass matrix $M^{}_{\rm D} \equiv v Y^{}_\nu/ \sqrt{2}$, and the effective Majorana neutrino mass matrix $M^{}_\nu \equiv - M^{}_{\rm D} M^{-1} M^{\rm T}_{\rm D}$, where the neutrino mass matrix is generated by the tree-level Weinberg operator. Second, as in the SM, $M^{}_W = g^{}_2 v/2$ is the $W$-boson mass, $\theta^{}_{\rm w} = \arctan \left(g^{}_1/g^{}_2 \right) $ is the weak mixing angle, and $F^{}_{\mu\nu} = \partial^{}_\mu A^{}_\nu - \partial_\nu A^{}_\mu $ is the gauge field strength tensor with $A^{}_\mu$ being the photon field.

The kinetic terms, lepton mass terms, the charged- and neutral-current interactions in Eq.~(\ref{eq:Lssb}) appear at tree level while the remaining operators at one-loop level. As before, we shall work in the flavor basis where both $M^{}_l = {\rm Diag} \{ m^{}_e, m^{}_\mu, m^{}_\tau \}$ and $M = {\rm Diag} \{ M^{}_1, M^{}_2, M^{}_3 \}$ are diagonal. One can first make a transformation $\nu^{}_{\rm L} \to V \nu^{}_{\rm L}$ with $ V = \bm{1} - R R^\dagger /2 $ and $R \equiv  M^{}_{\rm D} M^{-1}$ to normalize the kinetic term of left-handed neutrino fields, and then another one $\nu^{}_{\rm L} \to U^{}_0 \nu^{}_{\rm L}$ such that $U^\dagger_0 V M^{}_\nu V^T U^*_0 = \widehat{M}^{}_\nu = {\rm Diag}\{m^{}_1, m^{}_2, m^{}_3\}$. After these transformations, the effective Lagrangian in Eq.~(\ref{eq:Lssb}) becomes
\begin{eqnarray}\label{eq:Lsnd}
  \mathcal{L}^{}_{\rm SEFT} &=&  \overline{\nu^{}_{\rm L}} {\rm i} \slashed{\partial} \nu^{}_{\rm L} - \left( \overline{l^{}_{\rm L}} M^{}_l l^{}_{\rm R} + \frac{1}{2} \overline{\nu^{}_{\rm L}} \widehat{M}^{}_\nu \nu^{\rm c}_{\rm L} + {\rm h.c.} \right) + \left( \frac{g^{}_2}{\sqrt{2}} \overline{l^{}_{\rm L}} \gamma^\mu U \nu^{}_{\rm L} W^-_\mu + {\rm h.c.} \right)
  \nonumber
  \\
  && + \frac{g^{}_2}{2\cos \theta^{}_{\rm w}} \overline{\nu^{}_{\rm L}} \gamma^\mu U^\dagger U \nu^{}_{\rm L} Z^{}_\mu - \frac{e g^2_2}{12\left(4\pi\right)^2 M^2_W} \overline{l^{}_{\rm L}} \sigma^{}_{\mu\nu} R R^\dagger M^{}_l l^{}_{\rm R} F^{\mu\nu} + {\rm h.c.}
\;,
\end{eqnarray}
where the identities $e= g^{}_1 \cos \theta^{}_{\rm w} = g^{}_2 \sin \theta^{}_{\rm w}$ have been implemented, and the PMNS matrix $U = V U^{}_0$ is non-unitary as $U^{}_0$ is a unitary matrix but $V$ not. It is worthwhile to point out that the nonunitarity of $U$ and the flavor-changing neutral-current interaction are attributed to the existence of the dimension-six operator at the tree level in Eq.~(\ref{eq:6tree}).

\begin{figure}[t]
  \centering
  \includegraphics[width=1\textwidth]{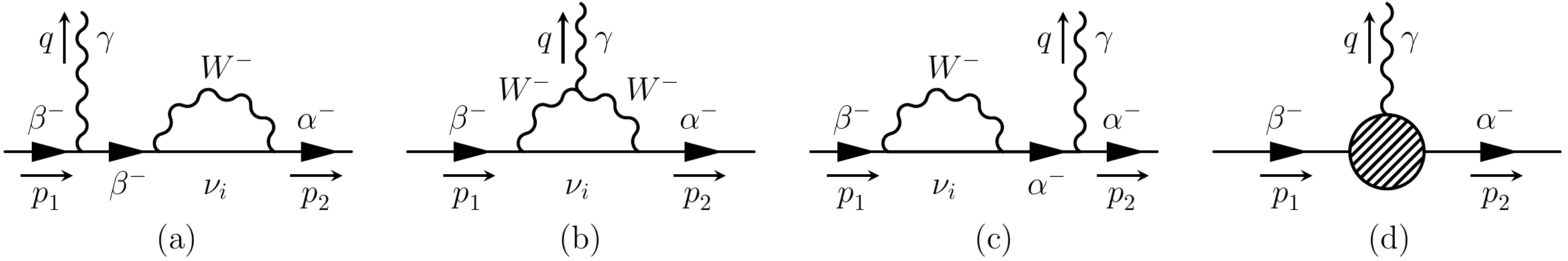}
  \vspace{-0.5cm}
  \caption{Feynman diagrams for radiative decays $\beta^- \to \alpha^- + \gamma$ decays at one-loop level in the {\it unitary} gauge. While (a)-(c) are mediated by massive neutrinos with a non-unitary flavor mixing matrix, (d) is generated by the dimension-six operators at one-loop level.}
  \label{fig:m2e}
\end{figure}

Thanks to nonzero neutrino masses and the nontrivial flavor mixing, the leptonic charged-current interaction induces radiative decays of charged leptons via the diagrams (a)-(c) shown in Fig.~\ref{fig:m2e}. The sum over all three amplitudes can be found in Refs.~\cite{Cheng:1980tp, Ilakovac:1994kj, Alonso:2012ji, Xing:2020ivm},\footnote{In this work, we focus on the non-supersymmetric canonical seesaw mechanism. We refer the reader to Refs.~\cite{Hisano:1995cp,Hisano:1995nq} for charged lepton flavor violation in the supersymmetric seesaw scenarios.} i.e.,
\begin{eqnarray}\label{eq:lfv1}
  {\rm i} \mathcal{M}^{}_{abc} = \frac{-{\rm i} e g^2_2}{2\left(4\pi\right)^2 M^2_W} \sum^{3}_{i=1} U^{}_{\alpha i} U^\ast_{\beta i} \left( - \frac{5}{6} + \frac{m^2_i}{4 M^2_W} \right) \left[ \epsilon^\ast_\mu \overline{u} \left(p^{}_2\right) {\rm i} \sigma^{\mu\nu} q^{}_\nu \left( m^{}_\alpha P^{}_{\rm L} + m^{}_\beta P^{}_{\rm R} \right) u \left(p^{}_1\right) \right] \; ,
\end{eqnarray}
where the repeated indices $\left( \alpha, \beta \right) = \left(e, \mu\right)$, $\left( e, \tau \right)$, $\left( \mu, \tau \right)$ characterize the final and initial charged leptons, and should not be summed over. The dimension-six operators at the one-loop level (i.e., those in the third line of Eq.~(\ref{eq:Lsnd})) contribute to radiative $\beta^- \to \alpha^- + \gamma$ decays via diagram (d) in Fig.~\ref{fig:m2e}. From the third line of Eq.~(\ref{eq:Lsnd}), it is straightforward to read off the corresponding amplitude of diagram (d)
\begin{eqnarray}\label{eq:lfv2}
  {\rm i} \mathcal{M}^{}_{d} = \frac{{\rm i} e g^2_2}{6\left(4\pi\right)^2 M^2_W} \left( R R^\dagger \right)^{}_{\alpha\beta} \left[  \epsilon^\ast_\mu \overline{u} \left(p^{}_2\right) {\rm i} \sigma^{\mu\nu} q^{}_\nu \left( m^{}_\alpha P^{}_{\rm L} + m^{}_\beta P^{}_{\rm R} \right) u\left(p^{}_1\right) \right] \;.
\end{eqnarray}
Therefore, the total amplitude for the radiative decays of charged leptons in the one-loop SEFT can be obtained by adding Eq.~(\ref{eq:lfv1}) and Eq.~(\ref{eq:lfv2}), i.e.,
\begin{eqnarray}\label{eq:lfv}
  {\rm i} \mathcal{M}^{}_{\rm tot} = {\rm i} \mathcal{M}^{}_{abc} + {\rm i} \mathcal{M}^{}_{d} &=& \frac{-{\rm i} e g^2_2}{2\left(4\pi\right)^2 M^2_W} \left[ \sum^{3}_{i=1} U^{}_{\alpha i} U^\ast_{\beta i} \left( - \frac{5}{6} + \frac{m^2_i}{4 M^2_W} \right) - \frac{1}{3} \left( R R^\dagger \right)^{}_{\alpha\beta} \right]
  \nonumber
  \\
  && \times \left[ \epsilon^\ast_\mu \overline{u} \left(p^{}_2\right) {\rm i} \sigma^{\mu\nu} q^{}_\nu \left( m^{}_\alpha P^{}_{\rm L} + m^{}_\beta P^{}_{\rm R} \right) u \left(p^{}_1\right) \right] \;.
\end{eqnarray}
One can easily verify that the $3\times 6$ matrix $(\begin{matrix} U & R \end{matrix})$ is approximately the upper $3\times6$ submatrix of the full $6\times6$ unitary matrix $\mathcal{U}$ that is used to diagonalize the overall $6\times6$ Majorana mass matrix of neutrinos in the type-I seesaw model, and $U U^\dagger + R R^\dagger = V V^\dagger + R R^\dagger = \bm{1} + \mathcal{O}\left( M^{-4} \right)$ holds. As expected, our result in Eq.~(\ref{eq:lfv}) is in perfect agreement with that obtained in the full theory~\cite{Cheng:1980tp, Ilakovac:1994kj, Alonso:2012ji, Xing:2020ivm} in the limit of $m^{}_i \ll M^{}_W \ll M^{}_j$ (for $i,j=1,2,3$), as recently demonstrated in Ref.~\cite{Xing:2020ivm}. It is now clear that compared to the corresponding result in the full theory, the total amplitude in the SEFT receives two equally important contributions, where one arises from light massive neutrinos (via the tree-level dimension-five operator) with non-unitary flavor mixing (via the tree-level dimension-six operator) and the other from one-loop dimension-six operators.

Starting with the amplitude in Eq.~(\ref{eq:lfv}), one can easily figure out the decay rate
\begin{eqnarray}\label{eq:dr}
  \Gamma \left( \beta^- \to \alpha^- + \gamma \right) \simeq \frac{\alpha^{}_{\rm em} G^2_{\rm F} m^5_\beta}{ 128 \pi^4} \left| \sum^{3}_{i=1} U^{}_{\alpha i} U^\ast_{\beta i} \left( - \frac{5}{6} + \frac{m^2_i}{4 M^2_W} \right) - \frac{1}{3} \left( R R^\dagger \right)^{}_{\alpha\beta}  \right|^2 \; ,
\end{eqnarray}
where $\alpha^{}_{\rm em} \equiv e^2/(4\pi)$ is the electromagnetic fine-structure constant, $G^{}_{\rm F}$ is the Fermi constant and the tiny ratio $m^2_\alpha/m^2_\beta \ll 1$ has been neglected. Thus the dimensionless ratio between the rate of radiative decays and that of the purely leptonic decays $\beta^- \to \alpha^- + \overline{\nu^{}_\alpha} + \nu^{}_\beta$ is given by
\begin{eqnarray}\label{eq:rat}
 \frac{\Gamma\left( \beta^- \to \alpha^- + \gamma \right)}{\Gamma\left( \beta^- \to \alpha^- + \overline{\nu^{}_{\alpha}} + \nu^{}_{\beta}\right) }
  &\simeq& \frac{3\alpha^{}_{\rm em}}{2\pi} \left| - \frac{5}{6} \left( U U^\dagger \right)^{}_{\alpha\beta} - \frac{1}{3} \left( R R^\dagger \right)^{}_{\alpha\beta}  \right|^2 =
\frac{3\alpha^{}_{\rm em}}{8\pi} \left| \left( R R^\dagger \right)^{}_{\alpha\beta}  \right|^2 \; ,
\end{eqnarray}
where higher-order terms have been omitted and the relation $U U^\dagger + R R^\dagger \simeq \bm{1}$ has been used in the last step. At this point, we clarify that the MUV scheme proposed in Ref.~\cite{Antusch:2006vwa} can be regarded a practically useful framework to constrain leptonic unitarity violation by experimental data from electroweak precision measurements, lepton-flavor-violating decays of charged leptons and neutrino oscillations without specifying the UV full theory. However, to test any UV full theory by precision data at low-energy scales, one has to construct the effective theory and determine the Wilson coefficients of higher-dimensional operators by performing tree- or loop-level matching in a consistent way. For radiative $\beta^- \to \alpha^- + \gamma$ decays, the results in the MUV scheme correspond to those in Eq.~(\ref{eq:lfv}) and Eq.~(\ref{eq:rat}) without the $-R R^\dagger/3$ term. In contrast, in the SEFT with one-loop matching, additional dimension-six operators come into play and contribute such a term, ensuring the consistency between the SEFT and the full theory.

\vspace{0.5cm}

\framebox{\bf 5} --- The solid experimental evidence for neutrino masses and lepton flavor mixing indicates that the standard model is actually incomplete and can only serve as a low-energy effective theory. Unfortunately, there has been so far no direct and significant hint for new physics in all terrestrial experiments other than neutrino oscillations. Almost for all kinds of particle physics experiments, the primary goal is to precisely measure fundamental parameters in nature and search for possible deviations from the SM predictions, which signify the existence of new physics. In this regard, effective theories have proved to be a very powerful and convenient tool. Therefore, the theoretical foundations for constructing effective theories and their applications to concrete problems have received tremendous attention recently. This is also the case for neutrino physics.

In this work, we take one of lepton-flavor-violating processes, i.e., radiative decays of charged leptons, in the type-I seesaw model as an example to illustrate how to derive the low-energy seesaw effective field theory with one-loop matching. It has been explicitly shown that the one-loop processes $\beta^- \to \alpha^- + \gamma$ should be studied by using the seesaw effective theory with one-loop matching. All relevant dimension-six operators and the associated Wilson coefficients are derived. However, it should be noticed that the matching has been carried out at the high-energy scale and the renormalization-group running of the Wilson coefficients has been ignored. In principle, one should follow the standard procedure to decouple heavy particles of similar masses at one energy scale and perform the matching at this decoupling scale. Once the effective theory below the decoupling scale is constructed, the renormalization-group equations will be implemented to evolve all physical parameters to the next scale of heavy particles. Such a procedure should be repeated until the scale of relevant experiments is reached. We leave such a complete construction of the seesaw effective theory for future works.

\vspace{0.3cm}
{\sl This work was supported in part by the National Natural Science Foundation of China under grant No.~11775231, No.~11775232, No.~11835013 and No.~12075254, and by the CAS Center for Excellence in Particle Physics.}

\end{document}